\documentstyle[sprocl]{article}

\input{psfig}

\def\Journal#1#2#3#4{{#1} {\bf #2}, #3 (#4)}

\begin{document}

\title{OUR BEST FRIEND, THE COMA CLUSTER \\ (A HISTORICAL REVIEW)}

\author{Andrea BIVIANO}

\address{Istituto Te.S.R.E., \\
Area della Ricerca del CNR, via Gobetti 101, I-40129 BOLOGNA \\
E-mail: abiviano@iso.vilspa.esa.es}

\maketitle

\abstracts{ In this paper I describe
how our knowledge and understanding of the properties and structure of
the Coma cluster of galaxies has evolved through the years, since early
this century, when the first maps of the density of nebul\ae~ 
in the Coma region 
were produced, and until the present time. 
It is shown that most of the recent discoveries that have led to a change
in our view of this cluster, were in fact anticipated very early on.}

\section{Introduction: Why Coma?}\label{s-why}
Coma is one of the most studied clusters of galaxies of the sky, along with
Virgo and Perseus\footnote{Of the three clusters, Perseus was never dedicated a
whole conference, while it was the case for Virgo\cite{virgo} and Coma
(these proceedings).}.
While being the most distant of the three, 
with a mean redshift $z \simeq 0.23$, Coma has also been
the most appealing to observers
because of its location near the galactic pole (bII$ = 88^{\circ}$) and because
of its richness\footnote{Of the nearby clusters (distance class $\leq 1$) only
five are as rich or richer than Coma (according to Abell et al.\cite{aco}).}.

Another characteristic that differentiates Coma from Virgo and Perseus is
its regular and (roughly) spherical shape.
In Shane \& Wirtanen\cite{sw}'s words:
\begin{quotation}
{\em There appear to be two extreme structural types among the populous
clusters, exemplified by the Virgo and the Coma clusters. The Virgo type is
characterized by the absence of a strong central condensation and by lack of
symmetry [\ldots] The Coma-type cluster is characterized by a strong central
condensation and a tendency towards spherical symmetry.}
\end{quotation}

What could be more charming than spherical symmetry (even if only approximate,
see e.g. Schipper \& King\cite{sk}) for a theoretician? It is indeed not 
surprising that Coma has been chosen as the prototype cluster by theoreticians,
since the early papers
of Zwicky\cite{zw1}$^,$\cite{zw2} and others (e.g. Carpenter\cite{car},
Holmberg\cite{hol}, Tuberg\cite{tub}), 
and until the recent estimates of the density
of matter in the Universe (e.g. the "baryon catastrophe" of 
Briel et al.\cite{bhb}) and recent scenarios for the structure formation
(e.g. the "filament" scenario of West et al.\cite{wjf}).

Nevertheless, Coma's richness and regularity are not typical of all clusters.
As Kent \& Gunn\cite{kg} pointed out:
\begin{quotation}
{\em Coma is quite atypical among clusters in its richness,
compactness, and degree of symmetry.}
\end{quotation}
On the other hand, the title of this conference indicates that we now have
"a new vision" of Coma, that emerged through the 
(once controversial) works of many people
(e.g. Fitchett \& Webster\cite{fw}, Mellier et al.\cite{m88}).
As Biviano et al.\cite{b96} wrote:
\begin{quotation}
{\em Coma can now be considered as the prototype of rich clusters
endowed with subclusters, and thus not fully relaxed}
\end{quotation}

\begin{figure}
\centerline{\psfig{figure=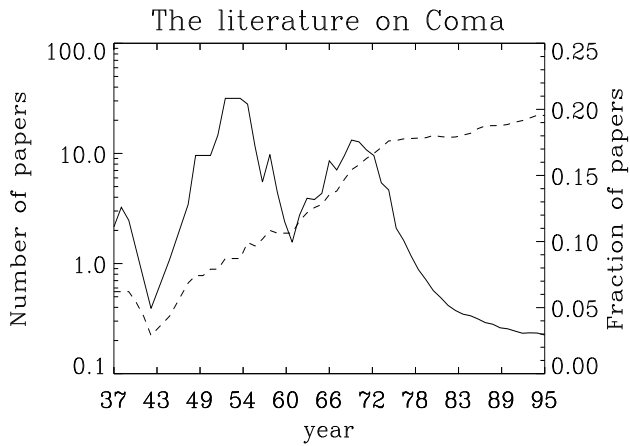,width=8cm,height=6.2cm}}
\caption{The number of papers published on the Coma cluster
through the years (dashed line), and the ratio between the number of
papers published on Coma and the 
number of papers published on galaxy clusters in general (solid line).}
\label{f-comapub}
\end{figure}

The interest of the astronomical community in Coma can be traced by counting
the number of publications dealing with this cluster, through the years.
While the total number of publications on Coma
has continuously increased with time (Fig.\ref{f-comapub}), 
the increase rate diminished in the 70's, when 
Coma became one of many well-studied clusters. This can be also seen by
plotting the same number divided by the
number of publications on galaxy clusters in general
(Fig.\ref{f-comapub}): there is a
continuously decreasing trend from the 70's on. No
doubt the proceedings of this conference will mark a change in this trend!

In the next sections I will describe how our vision of Coma has evolved
through the years\footnote{I have stopped my historical review with the year
1995. More recent works are cited only occasionally (with a possible
bias towards my owns!).}. For topics not covered in this review, I refer
the reader to the contributions of
Feretti, Gavazzi, Jones, and West in these same proceedings.

\section{Coma: an Old Friend}\label{s-old}
\subsection{The Myth}
The origin of the name "Coma Berenices" dates back to
the year $\sim 245$~B.C, when Ptolemy~III, the Egypt pharaon, left
his country to make war against Syria. His wife, Berenices, 
worried for her husband's safety, offered a lock of her
hairs to the goddess Arsin{\oe}s Zephiritis in the temple of
Can\={o}pus (near today's Ab\={u}q\={\i}r).

The lock misteriously disappeared during the night, and
princess Berenices felt very sad about what she considered a bad omen.
Conon, the court astronomer, told the princess
that the lock had been transformed into a star
constellation, {\em Coma Berenices,} i.e. Berenices' hairs. 

Apparently, the goddess appreciated Berenices' offer, infact Ptolemy came 
back safe.
Berenices had a pleasant and rich life until she was killed by one of her sons.

It was not until 1629 that the constellation name was used again, by Kepler. 

\subsection{The Coma cluster of {\em nebul\ae}}\label{s-nebulae}
\begin{figure}
\centerline{\psfig{figure=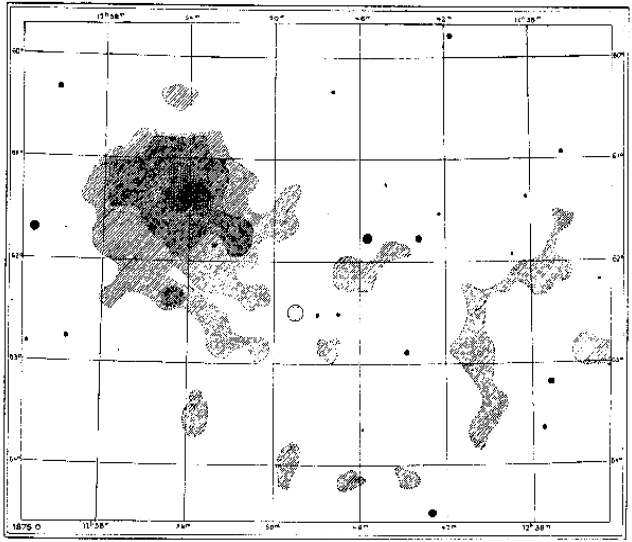,width=9cm,height=7.4cm}}
\caption{The density of nebulae in the region of Coma, according to
Wolf. Note the south-western extension (north is up, east 
is to the left). Every grid element is 28'$\times$60'.}
\label{f-wolf}
\end{figure}

In historic times, Herschel\cite{her}
was the first to notice the concentration of
nebul\ae~ in the constellation of Coma Berenices. A more rigorous catalogation
of nebul\ae~ in the Coma region was done by Wolf\cite{wo1}$^,$\cite{wo2} 
in the early years of this century (following up an earlier work of
D'Arrest\cite{dar}). Wolf counted 108 nebul\ae~ in a circle of 30' diameter.
The number of catalogued nebul\ae~ in the Coma region increased to more than
300 with the observations of Curtis\cite{cur}. Wolf's 
map of the density of nebul\ae~ in the Coma region (see Fig.\ref{f-wolf}), 
already shows the elongated shape of the cluster in the south-west direction,
where a secondary density concentration appears to lie (the south-west
subcluster, see \S~\ref{ss-sw}).

Hubble \& Humason\cite{hh} first measured the velocities of a few cluster 
nebul\ae, ranging from 5100~km/s to 8500~km/s (an interval quite close to 
the real full extension of Coma in velocity space -- see, e.g. Gavazzi
et al.\cite{ga95}).

\section{Weighing Coma}\label{s-mass}
\subsection{Zwicky's Heresy}\label{ss-zwicky}
The mass of Coma was first estimated 
by Zwicky\cite{zw0}$^,$\cite{zw1} 
to be\footnote{Throughout this paper I use $H_0 = 50$~km/s/Mpc, and scale all
$H_0$-dependent quantities accordingly. Note
that the recent measurement of the Sunyaev-Zel'dovich effect in Coma (Herbig et
al.\cite{hlrg}) implies $H_0=71^{+30}_{-25}$~km/s/Mpc, consistent with the
value adopted here.}
$M > 5 \times 10^{14} M_{\odot}$, using the virial theorem.
This estimation was based on a value\cite{zw3} of 1200~km/s for the radial 
velocity dispersion of the cluster galaxies, $\sigma_v$,
not too far from current estimates (e.g. Colless \& Dunn\cite{cd}).

The corresponding mass-to-light ratio was 
large, $M/L > 50 M_{\odot}/L_{\odot}$, and a form of unvisible matter seemed
needed. Zwicky suggested that this dark matter could be detected as diffuse
IC light.

Zwicky's hypothesis of some form of dark matter dominating the cluster
dynamics, was not accepted by his contemporaries. 
Holmberg\cite{hol} considered it {\em "an unlikely assumption",} and
his scepticism was still shared by the Burbidges\cite{bur} and
de~Vaucouleurs\cite{dev} 20 years later! However, the alternative 
hypothesis, clusters being unbound and expanding systems, would imply
a very short timescale for disruption. This was found to be incompatible
with the large number of galaxy clusters in the sky,
and the similarity of nearby and distant ($z \simeq 0.2$) clusters
(Zwicky\cite{zw4}, Limber\cite{lim}).

Had Zwicky grossly overestimated the total cluster mass?
Zwicky\cite{zw3} himself pointed out that the application of the
virial theorem may be of only limited validity when the system has an irregular
distribution of galaxies, implicitely questioning the results obtained by
Smith\cite{smi} on the Virgo cluster, and anticipating recent results on 
clusters affected by substructures. 
The problem of outliers in the velocity distribution was first considered
by Schwarschild\cite{schw}.
The lower limit
he set to the velocity dispersion of Coma, $\sigma_v > 630$~km/s, was still
too high to get rid of the dark matter problem.

\subsection{More Data!}
A step further in the understanding of the mass and structure of Coma was done
by Mayall\cite{may}, thanks to the new technology of electronic photography.
In Fig.3 of his paper
$\sim 50$ galaxy velocities are plotted vs. clustercentric distance, $d$, and  
the decrease of $\sigma_v$ with $d$ is already quite evident. 
Despite this significant progress, Mayall complained that:
\begin{quotation}
{\em \ldots it is doubtful that satisfactory answers will be obtained until
there are at least a hundred velocities available for discussion, and
several hundred would be much better. If this is the case, then the current
rate of less than 10 velocities per year is impracticably slow.}
\end{quotation}
It is ironic that the actual average rate since the 60's has been only twice 
as high\footnote{The total number of currently available velocities for Coma 
cluster galaxies is $\geq 800$ (see van~Haarlem's contribution in these 
proceedings), i.e. 750 new velocities in the last 37 years.}
as the "impracticably low" rate in Mayall's times!

\begin{figure}
\centerline{\psfig{figure=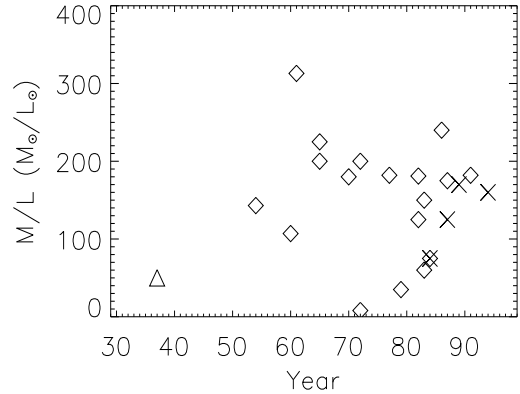,width=8cm,height=6.2cm}}
\caption{Several estimates of Coma M/L vs. the year when they were derived.
Diamonds represent estimates based on optical data, X's represent estimates
based on X-ray data; a triangle represent Zwicky's original lower limit 
estimate.}
\label{f-ml}
\end{figure}

The first numerical simulation of the evolution of
a Coma-like cluster (Peebles\cite{pee}) showed that the 3D-$\sigma_v$ 
should decrease with
increasing clustercentric distance. Nearly simultaneously, the decrease of
projected-$\sigma_v$ was actually measured by Rood\cite{ro70} in Coma. 
He pointed out 
that such a radial trend of $\sigma_v$ could be due to a real dependence
of the 3D-$\sigma_v$ with radius, or to an anisotropic distribution of galaxy
orbits. In the early 70's Coma M/L estimates were already quite close 
to current estimates (see Fig.\ref{f-ml}).

\begin{figure}
\centerline{\psfig{figure=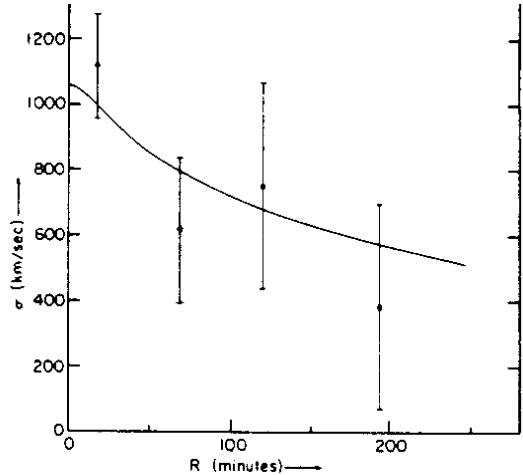,width=7.2cm,height=6.5cm}}
\caption{The radial velocity dispersion profile of Coma, km/s vs. arcminutes
-- from Rood et al.}
\label{f-rpkk}
\end{figure}

The density profile, accurately determined by Omer et al.\cite{opw} and
Rood et al.\cite{rpkk}, in combination with the $\sigma_v$-profile (see 
Fig.\ref{f-rpkk}), was used by Rood et al. to derive Coma's M/L,
and constrain the orbital anisotropy of Coma galaxies. They came to the
conclusion that the density and velocity dispersion profiles are 
{\em "consistent with an isotropic velocity distribution".}
Ivan King\cite{kin}, the last author in Rood et al.'s paper, relaxed
this conclusion. He noted that, in fact, several distributions 
of the galaxies and the dark matter were consistent with the data, 
and current cluster mass estimates could be systematically in error 
by a factor three. Ten years after, Kent \& Gunn and Bailey\cite{bai} 
arrived at (rouhgly) the same conclusions of Rood et al. and, 
respectively, King!

In the following years, mainly through the observations of 
Gregory \& Tifft\cite{gre}$^,$\cite{gt1}$^,$\cite{gt2}$^,$\cite{tg1}, the 
total number of measured redshifts for Coma galaxies increased to over 200.

\subsection{Beyond the Virial Theorem}\label{ss-beyond}
With some hundreds velocities available, more detailed models became possible.
A new heresy came around, the "binary model" of 
Valtonen \& Byrd\cite{vb1}$^,$\cite{vb2}, somewhat anticipated in the works
of Gainullina\cite{gai}, and Wesson et al.\cite{wl}$^,$\cite{wlg}. 
Valtonen \& Byrd suggested that Coma could be dymanically dominated by a
tightly bound couple of galaxies (NGC~4874 -- NGC~4889). Their model,
while reducing the {\em global} dark-matter discrepancy for the cluster,
nevertheless implied a very large M/L ratio for the 
two central dominant galaxies ($\sim 2000$). During the last 40 years,
Zwicky's heresy had become common sense; trying to reject the dark-matter
hypothesis had become the new heresy! This model was rejected 10 years
later, when The \& White\cite{tw1} showed that
Valtonen \& Byrd's model was 
inconsistent with the $\sigma_v$-profile at the centre of Coma.

\begin{figure}
\centerline{\psfig{figure=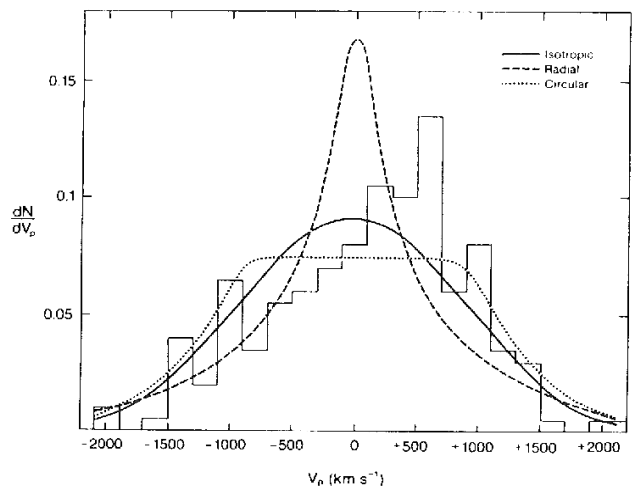,width=8.5cm,height=6.8cm}}
\caption{The observed line-of-sight distribution of the Coma galaxy 
velocities (in km/s)
with three theoretical models superposed -- from Merritt}
\label{f-mer}
\end{figure}

Other groups followed more traditional approaches.
Most (Kent \& Gunn, Millington \& Peach\cite{mp},
The \& White\cite{tw2}, Merritt\cite{mer}) came to the conclusions that the
best-fit model is also the simplest, i.e. light traces mass, and galaxy orbits
are isotropic throughout the cluster (Fuchs \&
Materne\cite{fm} disagreed, but their fitting method was found\cite{mp} 
to be very sensitive to the assumed form of the density profile).
While "simplest is best" provided an adequate description of the cluster
dynamics, other models could not be excluded, 
and the mass-to-light ratio of Coma was shown to be uncertain by a factor four
(from 50 to 200 $M_{\odot}/L_{\odot}$, see Bailey). 
Merritt first showed that the shape of the galaxy velocity
distribution at different radii contains information on the orbital
anisotropy. His attempt of fitting the velocity distribution of galaxies in 
Coma was unsuccesful though, because of the skewness of the 
velocity distribution (see Fig.\ref{f-mer}). It took 9 years to understand
that the skewness was caused by the contaimination of the SW group galaxy
velocities (Colless \& Dunn).

The X-ray observations of Gursky et al.\cite{gur} and Meekins et al.\cite{mee}
showed the existence of a hot IC gas component in Coma. The mass of this
new component was estimated by Gursky et al. 
to be quite below the total cluster mass, but far from negligible.

\begin{figure}
\centerline{\psfig{figure=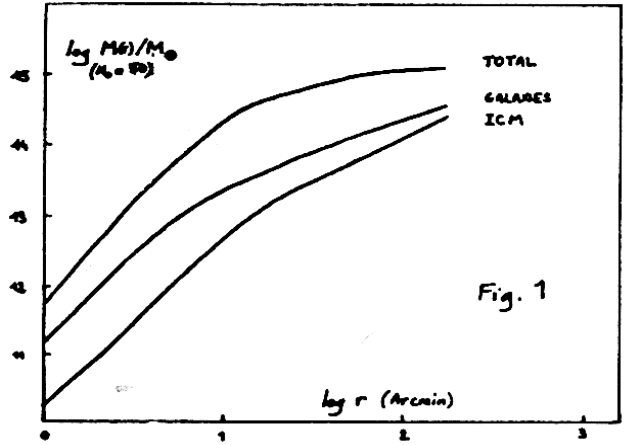,width=8.5cm,height=6.3cm}}
\caption{The mass (in solar mass units)
of the various components of the Coma cluster, as a function
of the clustercentric distance, in arcminutes, according to the model of
Gerbal et al. (log-log plot)}
\label{f-ger}
\end{figure}

The newly discovered IC gas mass component was taken into account in the
so-called "Multi-Mass Model" of Capelato,
Gerbal et al.\cite{ca1}$^,$\cite{ca2}$^,$\cite{for}$^,$\cite{ger},
where they also considered a spectrum of galaxy masses.
Capelato et al.\cite{ca2} suggested the existence of a virialized core
surrounded by a still collapsing halo.
Gerbal et al.\cite{ger} were possibly the first to show that the 
IC gas contribution to the total mass increases with the clustercentric 
distance (see Fig.\ref{f-ger}).

The basic uncertainty in the X-ray based mass estimation is the ignorance
of the detailed gas temperature profile. From the {\em HEAO 1 A-2} and 
{OSO 8} observations in the 2-60~keV band, Henriksen \& Mushotzky\cite{hm} 
deduced a steep gas temperature decrease with clustercentric distance. As a
consequence, they revised the total cluster mass estimate downward by
a factor four. Cowie et al.\cite{chm} reached a similar conclusion, by using
additional data from the {\em Einstein IPC.} 

\begin{figure}
\centerline{\psfig{figure=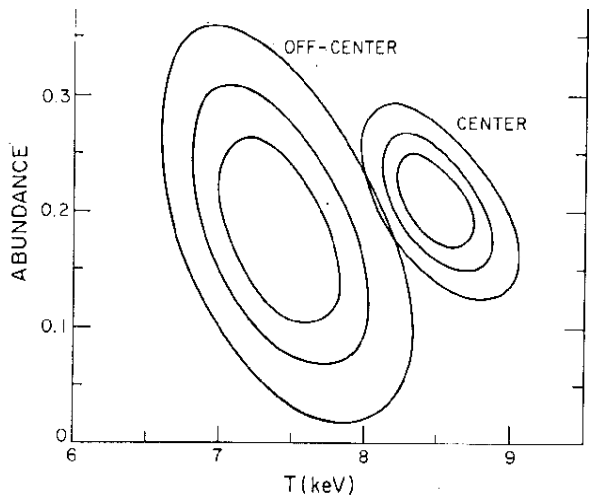,width=8cm,height=6.6cm}}
\caption{2D $\chi^2$ contours (at 68, 90 and 90 \% confidence level) for the
Iron abundance (with respect solar) vs. gas temperature (keV), 
for two pointings of {\em EXOSAT} - from
Hughes et al.}
\label{f-hughes}
\end{figure}

The \& White\cite{tw3} criticized these results, showing that such a
steep temperature decrease would require galaxies in the outer part of the
cluster to be on nearly circular orbits. Based on X-ray data, Hughes et
al.\cite{hug3}$^,$\cite{hug2}$^,$\cite{hug1} showed that the 
X-ray temperature does decrease off-centre (see Fig.\ref{f-hughes}),
but not as steeply as in Henriksen et al.'s model 
(which was shown to be internally inconsistent).

By combining the information from the optical and the
X-ray data, Hughes et al. constrained the total mass of
Coma within 1~Mpc to 3.9--7.2$\times 10^{14} M_{\odot}$
(1.1--3.0$\times 10^{15} M_{\odot}$ within 5~Mpc), and the mass-to-light ratio
to 90--250 $M_{\odot}/L_{\odot}$. Hughes also found 
that the ratio of the gas mass to the total mass 
increases with the distance from the cluster centre, 
thus confirming the early suggestion of Gerbal et al., and anticipating
Watt et al.\cite{wat}. 

Following works have not substantially modified our knowledge of the total mass
and mass-to-light ratio of Coma (current estimates are
$M \sim 2 \times 10^{15} M_{\odot}$ and
$M/L \sim 160 M_{\odot}/L_{\odot}$, see Fusco-Femiano \& Hughes\cite{ffh},
Makino\cite{mak}, and Hughes' contribution in these proceedings). 
The new measures of the IC gas temperature 
(see Briel and Honda in these proceedings) should allow to reach even better
accuracies.

\section{More Light on Coma}\label{s-lum}
The "missing mass" problem has always rather been a "missing light" problem.
Here I review the progress done in the estimation of the luminosity function
(hereafter LF) of Coma in the {\em optical.} 
Coma's LF has also been determined in several other bands, in the
radio (e.g. Willson\cite{wil}, Gavazzi et al.\cite{ga84}, 
Venturi et al.\cite{vgf}), in the IR (Gavazzi et al.\cite{ga95}), in 
the UV (Donas et al.\cite{dml1}). Anyway, the determination of these
non-optical LFs is quite recent, so I will not consider them in my
review. I refer the reader to the contributions of De~Propris, 
Gavazzi and Mobasher in these proceedings.

\begin{figure}
\centerline{\psfig{figure=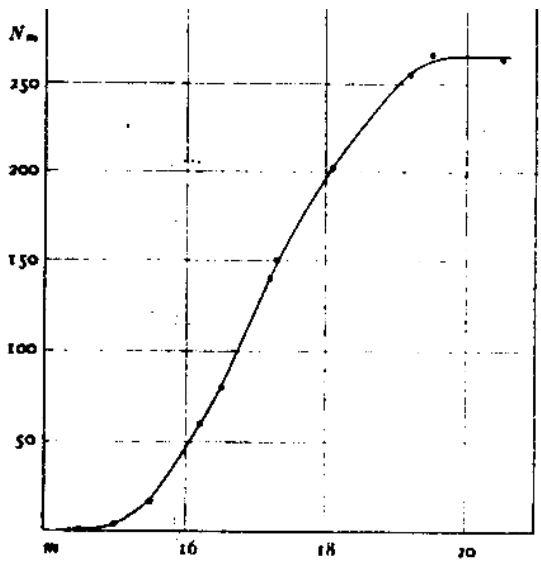,width=7.5cm,height=7.5cm}}
\caption{Cumulative counts of nebulae in the central region of Coma vs.
the photographic magnitude. In this figure, magnitudes range from 14 to
slightly more than 20, and counts from 0 to slightly more than 250
-- from Hubble \& Humason}
\label{f-hubble}
\end{figure}

\begin{figure}
\centerline{\psfig{figure=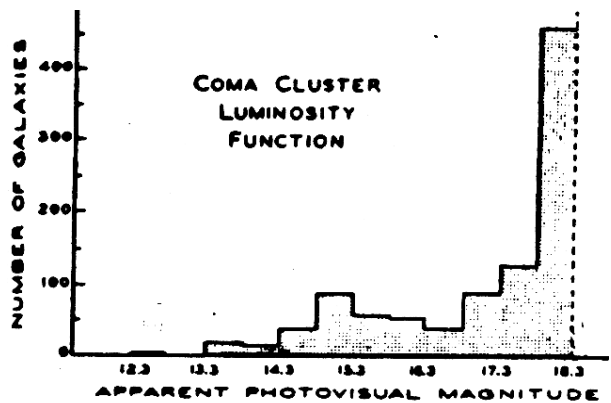,width=8cm,height=5.3cm}}
\caption{Differential LF of Coma galaxies. Magnitudes (on the x-axis)
range from  12.3 to 18.3, counts (on the y-axis) range from 0 to 500
-- from {\em Sky \& Telescope,}
{\bf XVIII,} 495 (1959)}
\label{f-abell}
\end{figure}

The first determination of the Coma LF was done by Hubble \& Humason.
The LF was found to peak at m$_{ph} \simeq
17$ (see Fig.\ref{f-hubble}). Twenty years later, Zwicky\cite{zw7} found
instead
\begin{quotation}
{\em \ldots that the luminosity function of the Coma cluster galaxies is
monotonely increasing with decreasing brightness}
\end{quotation}
and it took eight years more for Abell\cite{abe2} to show that 
in fact, both Hubble and
Zwicky were right, since the LF of Coma has a secondary maximum at bright
magnitudes (see Fig.\ref{f-abell}). 

The complex form of Coma LF was known since 1959, yet
since Schechter\cite{sch} proposed his analytical form for the
cluster LF, any bump or irregularity in the observed LFs tended to be 
overlooked\footnote{This is a further proof of the skill of astronomers 
in fitting a straight line to a circle!}. 
Many re-discoveries of this secondary maximum in the LF were needed before
astronomers start recognizing that some cluster
LFs are {\em not} well fitted by Schechter\cite{sch}'s function.

The first one to re-discover (or confirm) the secondary maximum
was Rood\cite{ro69}, 10 years later; then followed Godwin \& Peach\cite{gp},
in the late 70's, Thompson \& Gregory\cite{thg} in 1993, and finally
Biviano et al.\cite{b95} who closed this issue, by using only
spectroscopically confirmed cluster members from a complete galaxy sample.

New data allowed Rood\cite{ro69} to show that the LF is not the same for 
galaxies in the inner and outer regions of Coma. The secondary maximum 
looked more pronounced in the LF of galaxies in the inner region, where
the fraction of bright galaxies is higher. This result
was confirmed by Lugger\cite{lu86}$^,$\cite{lu87}, 10 years after
(despite the opposite conclusions reached by Gregory \& Tifft\cite{gt2}).

Numerical simulations allowed White\cite{wh76} to
show that dynamical friction can lead to the segregation of brighter
galaxies in the cluster centre, and a modification of the inner LF.
New simulations of Roos \& Aarseth\cite{ra} showed that also merging can
significantly affect the LF of galaxies in the centre of Coma.

The increase of Coma's LF at faint magnitudes seemed not to be strong enough to
provide all the "missing light"; galaxies fainter than m$_{pv}=18.3$ only
contribute $\sim 13$~\% of the total cluster light (Abell\cite{abe1}).
However, deeper observations were to indicate a further steepening of the LF.
First, Abell\cite{abe3} showed that
at magnitudes fainter than than m$_v=17.5$, the LF had an asymptotic
slope\footnote{$\alpha$ is defined by: 
$N(L) \, dL \propto L^{\alpha} \, dL$} $\alpha=-1.4$. Then, using
Godwin et al.\cite{gmp}'s new catalogue, 
Metcalfe\cite{met} determined an even steeper
slope ($\alpha=-1.9$) for the LF at $b \geq 19.74$. With such a steep
slope, the faint galaxy contribution to the total cluster light is
significant, $\sim 20$~\%. Metcalfe's result anticipated recent findings by
Lobo et al.\cite{lob} (who find $\alpha=-1.8$; 
see also the contributions of Adami, Lobo, and Sekiguchi in these proceedings).
Other evidences for a (more or less) steep LF at the faint end came from
Karachentsev et al.\cite{kkrv} and Bernstein et al.\cite{ber}.

\section{Looking into the Dark}\label{s-dark}
Observations and modelling of the Coma cluster have led theoreticians to
propose and eventually discard hypotheses on the nature of dark matter.
Here I review some of these hypotheses, although most are already ruled out,
and maybe none will turn out to be correct.

\subsection{No Dark Matter?}
Zwicky's original estimate of the large mass of Coma was regarded with
considerable scepticism at the beginning (see \S~\ref{ss-zwicky}). Possible
solutions to the problem of dark matter were that Coma (and clusters in general)
are unbound and expanding, or that the $\sigma_v$-estimate was boosted by the
presence of interlopers. In the words of Holmberg:
\begin{quotation}
{\em The temporary members with their hyperbolic velocities seem to offer a
more plausible solution of the difficulty.}
\end{quotation}

This solution to the problem was never totally discarded, since unidentified
subclustering is known to significantly affect cluster $\sigma_v$-estimates.
In the case of Coma, however, Schwarschild pointed out that
interlopers cannot led to more than a factor two overestimate of the cluster
$\sigma_v$. On the other hand, the instability argument could
not explain why we see so many clusters, nor the apparent lack of
significant evolution in the structure of clusters at different redshifts
(see \S~\ref{ss-zwicky}). 

The existence of dark matter was again questioned by Tifft\cite{tif}. He
discovered a correlation between galaxy redshifts and magnitudes that led him
to question the usual physical interpretation of a galaxy redshift. 
Tifft estimated that the intrinsic
Doppler velocity dispersion could be less than 220~km/s.
To my knowledge, this so-called "band-effect" has never really been
ruled out (it was even recently confirmed by Nanni et al.\cite{nan}), unless
Simkin\cite{sim} was right and the effect is an
artefact due to night-sky distorsion in the observed spectra.

Another viable solution to the "missing mass" problem, that does not require
any dark matter, came from Milgrom\cite{mil} who proposed a
modification of the theory of Newton dynamics. Nevertheless, MOND, this
new theory, cannot at the same time explain Coma dynamics and the spiral
rotation curves (The \& White\cite{tw4}).

\subsection{Diffuse Light}\label{ss-light}
Zwicky\cite{zw5}'s original approach to the "missing mass" problem 
consisted in looking 
for the missing light. He was the first to claim detection of
large luminous patches in the centre of Coma. Later on,
Welch \& Sastry\cite{ws}, Kormendy \& Bahcall\cite{kb}, Thuan \&
Kormendy\cite{tk} and Mattila\cite{mat} also found evidence for diffuse light
in the centre of Coma. The contribution of this diffuse component to the total
cluster mass was however estimated to be negligible (at most 3~\% of the total
mass, according to Kormendy \& Bahcall).

Mattila proposed several possible origins for this diffuse light:
extended galaxy envelopes, dwarf galaxies (see \S~\ref{s-lum}), globular
clusters, intergalactic stars, scattering by dust grains (see
\S~\ref{ss-dust}).

Recently, based on deep CCD observations,
Bernstein et al. showed that the diffuse light in Coma follows the
same distribution of globular clusters and dwarf galaxies, and 
the units that make up such a diffuse luminosity must be 
$\leq 10^3 L_{\odot}$.

\subsection{Diffuse Gas}\label{ss-gas}
As early as in 1956, the first detection of IC gas was claimed by
Heeschen\cite{hee}, based on 21-cm line-emission observations. Heeschen
concluded that the total mass in HI was $\sim 1/4$ of the total cluster mass.
However, his result was shown to be spurious by Muller\cite{mul}. 

Boldt et al.\cite{bmrs} first claimed detection of extended 
X-ray emission in the direction of
Coma, but this finding was shown to be inconsistent with
other observations (Friedman \& Byram\cite{fb}).

\begin{figure}
\centerline{\psfig{figure=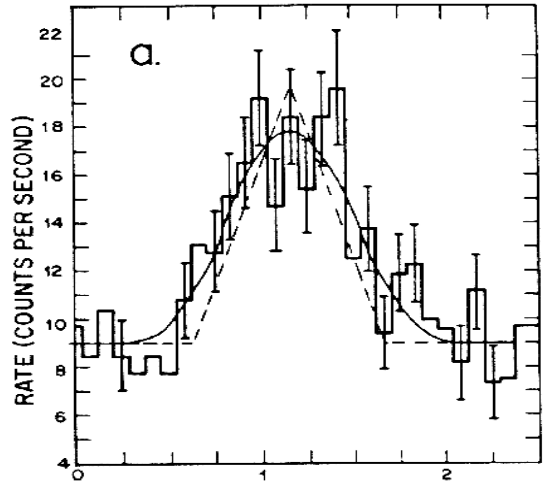,width=7.5cm,height=6.6cm}}
\caption{The counting-rate distribution vs. the relative azimuth 
(in degrees) obtained with
the {\em Uhuru} instrument in the direction of Coma; the solid line shows a fit
to the data with an extended source model, while the dashed line shows the
expected distribution for a point source -- from Gursky et al.}
\label{f-gursky}
\end{figure}

In 1970 Turnrose \& Rood\cite{tr} used the available H$\beta$ and X-ray data
to set an upper limit of $10^5$~K to the temperature of a diffuse gas with the
mass needed to bind the cluster. Their
work was published shortly before the real detection of 
Coma in the X-ray by Meekins et al. and Gursky et al. (see Fig.\ref{f-gursky}).
Assuming that the emission was thermal bremsstrahlung from an IC gas, 
Gursky et al. estimated its mass in a few percent only 
of the total cluster mass.

The detection of the Fe line with {\em OSO-8} by Serlemitsos et
al.\cite{ser}, proved the thermal nature of the X-ray emission. 
At the same time, the presence of metals in the IC gas indicated
that this had been at least partly processed in stars. 

Currently, the contribution of the hot gas to the total mass is 
known to increase
with the distance from the cluster centre, possibly up to
$\sim 50$~\% (Hughes). A cool diffuse gas component may have now been detected
(Lieu, these proceedings).

\subsection{Diffuse Dust}\label{ss-dust}
Zwicky\cite{zw6} noted a deficiency of clusters behind the Coma cluster
as compared to other regions of the sky. He interpreted this deficiency
as evidence
for IC extinction, presumably due to diffuse dust. Later, Noonan\cite{no1} 
found evidence for IC extinction of $\sim 0.4$ mag in the blue band, in
agreement with the estimate given by Karachentsev \& Lipovetskii\cite{kl}.
Few years later, Wesson\cite{wes} made the rather extreme
hypothesis that IC dust may be present in such large quantities as to bind the
Coma cluster!

The IC dust hypothesis encountered more criticism than consensus.
In the 60's de~Vaucouleurs noted that the fact that diffuse 
gas remained undetected implied a low density of diffuse dust, and 
Abell\cite{abe1} considered the evidence for IC inconclusive.
Smart\cite{sma} showed that IC dust must be significantly depleted because of
the dust grain sputtering by the hot IC gas. 
Tifft \& Gregory\cite{tg1} identified a group in the background of Coma and
showed that the magnitudes of galaxies in this group 
are not significantly affected by extinction.

Recently, Dwek et al.\cite{drm} modelled the formation and evolution of IC dust
in Coma, including sputtering from the IC gas and dust injection from galaxies.
They derived a dust density much below that required to explain the observed
visual extinction, but consistent with the upper limit reported by 
{\em IRAS} for IR emission in the Coma region. Ferguson\cite{fer}, using
the Mg$_{2}$--(B-V) relation for Coma ellipticals, set an upper limit
of E(B-V)~$\sim$~0.05 for IC extinction.

The amount of IC dust recently detected with {\em ISO} 
(Stickel, these proceedings) is even lower than what predicted by Dwek et
al.'s model.

\subsection{Galaxy Halos}
Ostriker \& Peebles\cite{op} suggested that the existence of massive halos
around spiral galaxies was needed in order for the disks 
to be stable against bar formation. Their paper led
Lecar\cite{lec} to suggest that the diffuse dark matter in
Coma comes from tidally torn-off halos. Indeed, the lack of significant
luminosity segregation of Coma galaxies may indicate that they have lost their
massive halos very early in the history of the cluster (see \S~\ref{ss-lseg}).

Support to his model came from the
observations of Thompson\cite{tho}, who showed
that the density of barred galaxies is higher in
the cluster centre, as predicted if these galaxies have
indeed lost their halos.

Lecar's hypothesis would imply similar
mass-to-light ratios for galaxies and clusters, and this is consistent with
current estimates (e.g. Bahcall\cite{bah1}).

\subsection{Particles}
Most today's cosmologists think that dark matter is made of 
some sort of weakly interacting massive particles filling the Universe.
Massive neutrinos have long been considered as possible candidates.
Cowsik \& McClelland\cite{cmc} were
the first to draw the attention of the astrophysical community to massive 
neutrinos. Their work is of relevance here, as they compared their model
to the observations of Coma. In their acknowledgments, we can read:
\begin{quotation}
{\em Our interest in the problem of the Coma cluster started after listening to
the excellent seminars on the subject by Professors Ivan R. King, Eugene D.
Commins, and Joseph Silk.}
\end{quotation}

\section{The Tidy Coma: A Place for Each Galaxy}\label{s-seg}
When we say that Coma is a cluster of galaxies, we better specify the kind of
galaxies we are speaking of. In fact, galaxies of different type
and luminosity have different distributions in the Coma cluster, and the
cluster looks different when only, say, the ellipticals, or the spirals are
selected. Here I review the history of the discovery of galaxy 
segregation in the Coma cluster.

\subsection{Luminosity segregation}\label{ss-lseg}
\begin{figure}
\centerline{\psfig{figure=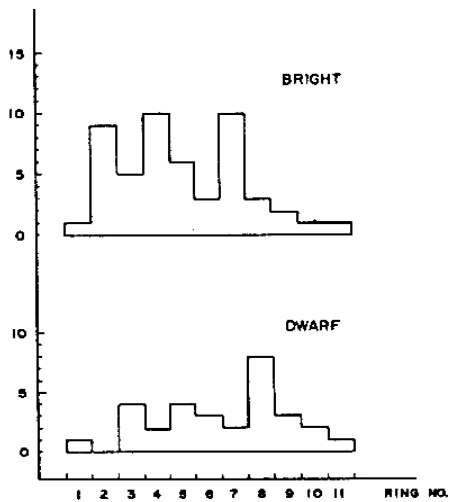,width=6cm,height=7cm}}
\caption{Bright and dwarf galaxy distributions in concentric rings centered on
the Coma cluster centre. The ring no. increases with clustercentric distance
-- from Rood \& Turnrose}
\label{f-rood}
\end{figure}

Zwicky\cite{zw4} was the first to notice that bright and faint
galaxies in Coma have different radial distribution, bright galaxies being more
concentrated. His finding was at odd with the conclusion reached by
Omer\cite{om52} few years before. In the 60's the evidence for luminosity
segregation had less supporters (Reaves\cite{rea}) than opponents
(Abell\cite{abe1} and Omer\cite{om66}), until the new data-set of 
Rood allowed Rood \& Turnrose\cite{rt} to conclude that dwarfs 
are indeed less concentrated than bright galaxies (see Fig.\ref{f-rood}).
Their results were confirmed by Noonan\cite{no1}.

However, the luminosity segregation in Coma is not a very strong effect.
Rood\cite{ro65} and White\cite{wh76}$^,$\cite{wh77} noted that a much
stronger effect would result from two-body relaxation if all the cluster 
dark mass was in galaxy halos. 

Using the new data-set of Godwin \& Peach, Capelato et al.\cite{ca1} showed
that luminosity segregation is a more complicated issue than previously
thought. While dynamical friction tends to produce a concentration of brighter
galaxies in the Coma centre, merging of these same galaxies tends to reduce the
number of intermediate brightness galaxies, and to create
ultra-bright galaxies. Observationally, this effect is seen as an 
anti-segregation of galaxies with m$_{25} \leq 14.5$. 
Capelato et al.'s observational evidence was reproduced in 
Roos \& Aarseth's numerical simulations only two years later
(see also Athanassoula, these proceedings).

Thompson \& Gregory showed that, among the dwarf galaxies, dwarf ellipticals
follow the same distribution of giant ellipticals, while dwarf
spheroidals are lacking in the core. This was interpreted as  evidence
for tidal disruption of the spheroidal galaxies (see also Secker, these
proceedings).

Recently, Biviano et al.\cite{b96} showed that faint galaxies
describe a more regular cluster than do bright galaxies, for which the effect of
subclustering is stronger, and interpreted this finding as evidence for
ongoing accretion of groups onto the cluster (see also
\S~\ref{s-slice} and Lobo, these proceedings).

\subsection{Spirals in Coma?}\label{ss-sp}
\begin{quotation}
{\em The fact that nebul\ae~ near the centre of concentrated clusters are
predominantly of the elliptic type, whereas spirals are relatively more
numerous on the outskirts of clusters \ldots}
\end{quotation}
\ldots was already well known at the time Zwicky\cite{zw2} was writing these 
lines, even if, in 1962, Neyman et al.\cite{nsz} maintained 
that most (if not all) of this effect was due to an
observational bias. Although
Andreon\cite{and1} has recently shown they were not
completely wrong, morphological segregation {\em is} real, and it can
be seen in Coma as in (almost) any other cluster (being particularly evident
when galaxies are selected in the UV, see Donas et al.\cite{dml1}).

In this respect, what distinguishes Coma from most other clusters, 
is the almost complete absence of spirals. Abell\cite{abe1}
maintained there are no spirals in Coma at all, in contrast with
Rood\cite{ro68} and Rood et al. who found that
some peculiar spirals do belong to Coma, and that
at least 16 of the spectroscopically confirmed members of Coma were spirals or
irregulars. Faced to the evidence that some spirals have velocities 
close to the mean cluster velocity, Abell\cite{abe3} 
made the hypothesis that these spirals
are members not of the cluster but of the Coma supercluster. 

Sullivan \& Johnson\cite{sj} observed three spirals in Coma and found that they
had a surprisingly low HI abundance for their luminosity, when compared to
similar spirals in the field, i.e. they were "HI-deficient".
The authors concluded that these spirals have passed
through Coma and have been stripped of part of their gas. 
Following studies (Sullivan et
al.\cite{sbbs}, Chincarini et al.\cite{cgh}, Bothun et al.\cite{bss},
Gavazzi et al.\cite{ga84}) not only confirmed these results,
but also showed that the HI-deficiency mostly concerns spirals
in the core of Coma, and not spirals in the Coma supercluster. This definitely
proved the existence of a population of cluster spirals (note however that
Coma spirals are {\em not} H$_2$-deficient, see Boselli, these
proceedings).

Doi et al.\cite{dfot}, via automatic
classification of galaxy types, have recently concluded
that the spiral fraction in Coma was previously underestimated (see
also the contribution of Andreon in these proceedings).

Note that, even if spirals are cluster members, dwarf irregulars are not
(Thompson \& Gregory).

\subsection{Velocity Segregation}\label{ss-vseg}
Different galaxies in Coma have different velocity distributions. This was
first noticed by Hawkins\cite{haw} who pointed out that galaxies in the Coma
centre have a lower mean velocity than galaxies at the edge. However, the
value he quoted for the central galaxies (6254~km/s) was very low, and 
has never been confirmed since.

The first firm result on the velocity segregation in Coma was obtained by
Rood et al. These authors noticed that the five brightest
galaxies (m$_p < 15.0$) of the Coma core have a very low velocity dispersion,
$\sigma_v=231$~km/s. Such an evidence was first confirmed by Struble\cite{str}
and more recently by Mellier et al. Struble considered two possible
explanations to this effect: (i) the existence of a subcluster in the Coma
core, and (ii) the result of dynamical friction (see also \S~\ref{ss-core}).

In the early 70's, Tifft and 
des~For\^ets \& Schneider\cite{dfs} noted that Coma ellipticals 
have a different mean velocity from non-ellipticals.
Ten years after, Kent \& Gunn showed that $\sigma_v$ increases along the Hubble
sequence, from ellipticals to S0s to spirals. This was recently confirmed in
the works of Andreon\cite{and2}, Biviano et al.\cite{b96}, and Colless \& Dunn.
Zabludoff \& Franx\cite{zf} compared the whole distributions (not only their
moments) of different morphological types, and found that 
the velocity distributions of ellipticals and spirals are different.

There is a general agreement in interpreting these results as evidence that the
blue/star-forming galaxies have not yet reached virial equilibrium, and are
still infalling into the cluster (some of them possibly for the first time).

\begin{figure}
\centerline{\psfig{figure=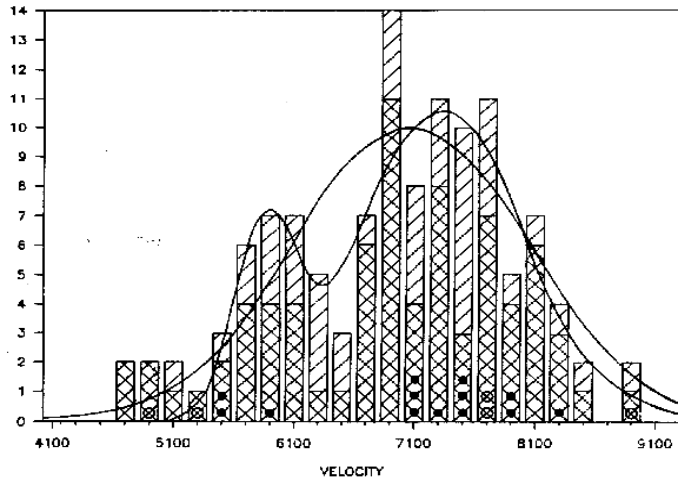,width=9cm,height=6.4cm}}
\caption{Histogram of velocities (in km/s)
within 1$^{\circ}$ of the centre of Coma
(shaded) and within 0.5$^{\circ}$ of Coma (double shaded); strongly
HI-deficient galaxies (filled circles), and moderate or null
HI-deficient galaxies (open circles), are shown. Curves represent Gaussian
fittings to the data -- from Gavazzi}
\label{f-gava}
\end{figure}

What remains possibly unexplained is the double-peaked 
velocity distribution characteristic of
many populations of Coma galaxies (see Fig.\ref{f-gava}):
\begin{itemize}
\item supercluster spirals (Gavazzi\cite{gav});
\item HI-deficient galaxies (Gavazzi\cite{gav});
\item post-starburst galaxies (Caldwell et al.\cite{cal}, see Fig.16 in Biviano
et al.\cite{b96});
\item UV-selected galaxies (Donas et al.\cite{dml2});
\item radio-galaxies (Kim et al.\cite{kkdl});
\item S0 galaxies (Zabludoff \& Franx);
\item blue galaxies (Biviano et al.\cite{b96}).
\end{itemize}

Of the two peaks, one is centered at $\sim 7500$~km/s (close 
to the mean velocity of the
SW group, see \S~\ref{ss-sw}), and 
the other is centered at $\sim 5500$~km/s,
while the cluster mean velocity is $\sim 6900$~km/s. 
Subclustering would seem the most obvious explanation for this complex velocity
distribution, yet galaxies with velocities close to these peaks are not 
spatially subclustered (see also Gerbal, these proceedings).

\section{Slicing Coma}\label{s-slice}
Coma has long been considered the prototype of well-relaxed, regular 
clusters (see \S~\ref{s-why}). Nevertheless, also the existence 
of substructures in Coma has long been known. This is shown in the next
sections.

\subsection{The SW Group}\label{ss-sw}
\begin{figure}
\centerline{\psfig{figure=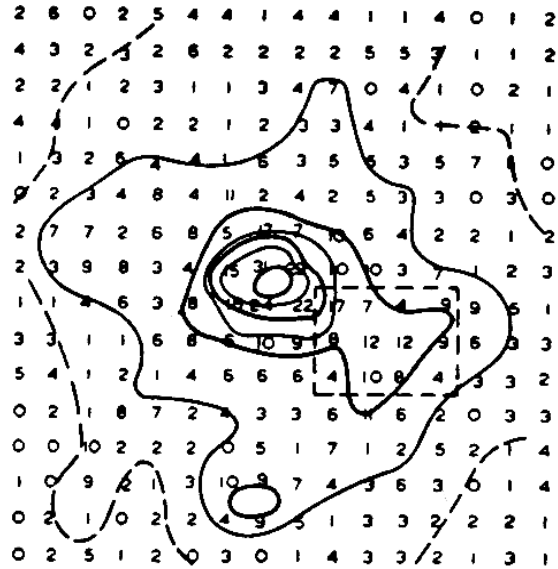,width=8cm,height=8cm}}
\caption{The density of nebul\ae~ in the region of Coma, according to
Shane \& Wirtanen. Note the south-western extension, indicated by dashed lines 
(north is up, east is to the left).}
\label{f-sw}
\end{figure}

The existence of a subcluster at $\sim 0.5^{\circ}$ South-West of the Coma
centre, around the bright galaxy NGC~4839,
was already noticeable in the map of the density of nebul\ae~ in the
Coma region (see Fig.\ref{f-wolf}), that Wolf made in 1902. The presence of
the SW galaxy concentration was confirmed by Shane \& Wirtanen's galaxy
counts, half a century later (see Fig.\ref{f-sw}). Here is how
Shane \& Wirtanen described the SW subcluster:
\begin{quotation}
{\em It is apparent that there is a subsidiary concentration of nebul\ae~
southwest of the cluster center. This grouping may be a secondary feature of
the cluster or it may represent an independent aggregation.}
\end{quotation}
Note that Wolf's map was not mentioned in Shane \& Wirtanen's paper.

Following up Shane \& Wirtanen's "discovery" of the SW subcluster,
van den Bergh\cite{vdb} developed the first objective method for the 
detection of subclustering. He measured the apparent galaxy separations in
Coma, before and after scrambling the galaxy position angles, while keeping
their clustercentric radial distances unchanged. He found that
the apparent galaxy separations were smaller in the real cluster than in the 
synthetic one, and concluded:
\begin{quotation}
{\em Taken at face value, this result implies that 
subclustering occurs in the Coma cluster.}
\end{quotation}
His result motivated Abell\cite{abe4}$^,$\cite{abe5} to perform his own 
analysis of substructure. He found evidence for subclustering in 5 out of the 7 
clusters examined, but not in Coma.

Omer et al. built the Coma density profile, based on
three independent galaxy counts in the Coma region, and found
a clear secondary peak at $0.5^{\circ}$ radial distance.

The apparent overall regularity of Coma led
Rood \& Turnrose to suggest, by analogy, that the SW concentration 
was rather a background group than a subcluster.

\begin{figure}
\centerline{\psfig{figure=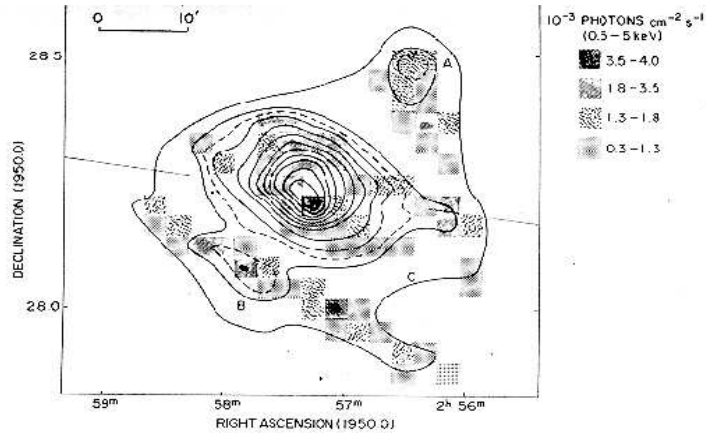,width=9.5cm,height=5.8cm}}
\caption{The map of the X-ray emission from 
Coma in the 0.5-5~keV band -- from Johnson et al.}
\label{f-johnson}
\end{figure}

In the 70's the first X-ray maps 
of Coma (Gorenstein et al.\cite{gfth} and Johnson et al.\cite{joh}) were
produced. The limited size of these first X-ray maps did not allow the
detection of the SW group. Nevertheless, a SW extension {\em was} visible in
the map of Johnson et al. (see Fig.\ref{f-johnson}).

In the 80's the SW group was finally identified also in the velocity space.
Baier\cite{ba1} reported\footnote{Sherbanowsky's paper is extremely difficult
(impossible?) to find even in very good libraries. It would 
have probably remained unnoticed, had Baier not mentioned it.}
on Sherbanowsky\cite{she}'s identification of a group
of 12 galaxies with an average velocity of 7437~km/s in the SW of Coma.
Sherbanowsky's estimate for the mean velocity of the SW group was
only $\sim 100$~km/s higher than the current estimate\footnote{Colless
\& Dunn estimate a mean velocity of the SW group of 7339~km/s and a
velocity dispersion of 329~km/s}.

Another (independent) detection of the SW group in the phase-space came from 
Perea et al.\cite{pmo} Their estimates of
the average velocity and $\sigma_v$ of the SW group, 
were however offset by 200~km/s from current ones\cite{cd}.

In spite of all previous evidences on the existence of a SW subcluster,
in 1988 Dressler \& Shectman\cite{ds}
claimed $\leq 6$~\% probability that Coma had substructures!
In the same year, however, Mellier et al. published their seminal paper, 
in which they identified {\em "not less than 9 local density peaks",} including
the SW group. Their estimates of the SW velocity moments
were nevertheless based on 4 galaxies only, and therefore rather uncertain.
Escalera et al.\cite{esm} put Mellier et al.'s results on a firm statistical
basis, using the wavelet method for structure detection. 

In the 90's the group around NGC~4839 was also discovered in the 
X-rays, first by Briel et al. using the 
{\em ROSAT PSPC,} and almost at the same time
by Watt et al. using the {\em SL2 XRT.} 

In the last years, the emphasis moved from the determination of the SW group
properties to the determination of its evolution in relation to Coma.
Has the group already passed through Coma, or is it infalling 
into the cluster for the first time? Both in the X-ray (White et al.\cite{wbh})
and in the radio (Cordey\cite{cor}) there is evidence
for a tail of gas behind NGC~4839, in the opposite direction to the cluster
centre, as if the galaxy was falling into the Coma core, and its gaseous
atmosphere was swept away by motion through an external medium. On the other
hand, Burns et al.\cite{brlk}'s numerical simulations suggest 
that the group has already passed through the cluster core\footnote{In their
paper Burns et al. note that {\em "the recession velocity of the NGC~4839 group
is slightly smaller (by $\sim 100$~km~s$^{-1}$) than that of the Coma cluster".}
This is wrong, the recession velocity of the group is $\sim 400$~km/s
{\em higher} than that of Coma.}.

The issue is still controversial (see Colless \& Dunn vs.
Biviano et al.\cite{b96}). Evidences in favour of the "first infall" scenario
are:
\begin{itemize}
\item the X-ray and radio tail of NGC~4839;
\item spiral galaxies in the SW group region do not show HI-deficiency
(Bravo, these proceedings);
\item the group properties (richness, velocity dispersion, X-ray luminosity),
are all consistent with a non-perturbed group (Colless, these proceedgins).
\end{itemize}
Nevertheless, this scenario cannot explain the presence of
a bridge of material connecting Coma and the group, detected
as diffuse optical light (Mattila), as a spray of post-starburst galaxies
(Caldwell et al., and Caldwell's contribution in 
these proceedings), in the X-ray (Briel et al.) and in the radio (Kim et
al.\cite{kkgv}). If the group has crossed the cluster, this bridge could
be interpreted as tidally stripped material from the group. The bridge is
unlikely to be related to the filament
connecting Coma to A1367 (see, e.g., West, these proceedings)
since the filament orientation is different from that of the bridge.
On the other hand, the North-East filament of the Coma supercluster has
the correct orientation. Several groups are found to lie in this filament,
and these are predicted to fall into the cluster in the future 
(West, these proceedings). 

This issue could be solved by an accurate determination of the relative
distances between Coma and the group, via the use of secondary distance
indicators (see the preliminary results of Lima-Neto in these proceedings).

\subsection{Subclusters in the Core}\label{ss-core}
The existence of substructure within the core of Coma was first suggested by
Bahcall\cite{bah2} in the 70's.
She observed an anisotropic distribution of the bright
galaxies in the Coma core, along the E-W direction and suggested that:
\begin{quotation}
{\em \ldots the observed anisotropy of the bright galaxies could arise from
subclustering around each of the two supergiants at the center of the cluster.}
\end{quotation}
She also noticed that subclustering was stronger around NGC~4874,
the less bright of the two central dominant galaxies. 
Recent results\cite{b96}$^,$\cite{cd} confirm her early
suggestions.

One year later, Rood\cite{ro74} pointed out {\em "a possible tendency for S0
galaxies to surround NGC~4874 and ellipticals to surround NGC~4889".} To my
knowledge, there has been no confirmation of this effect. The recent
morphological catalogue of Andreon et al.\cite{and3} is best suited for the
investigation of this issue (see also Andreon, these proceedings).

In January 1979 two papers appeared, both fundamental for the
understanding of subclustering in the Coma core, one from Quintana\cite{qui},
the other from Struble.

Quintana noticed a narrow density peak centered very close
to NGC~4874, and suggested it 
{\em "may indicate the presence of a dynamically separated subunit".}
Struble observed that the velocity dispersion of three clusters (among which
Coma) decreases when only the core region is selected. This was interpreted
as evidence that
\begin{quotation}
{\em \ldots the most massive galaxies in the core have "captured" less
massive galaxies as satellites and formed isolated subsystems}
\end{quotation}
Struble also mentioned dynamical friction as a possible alternative explanation
(see \S~\ref{ss-vseg}).

Valtonen \& Byrd's binary model for the Coma cluster (see
\S~\ref{ss-beyond}) was motivated by the growing evidence of the existence of
subclusters in the Coma core. This suggests
that by the late 70's the idea of a complex structure for the Coma core was
already taking root in the astrophysical community. In 1984, Baier published
the first paper specifically devoted to the issue of subclustering in Coma.
He reviewed the evidence for subclustering in Coma, partly coming from the
presence of the SW-group (see \S~\ref{ss-sw}), and partly from the {\em "double
character of the central cluster region".}

\begin{figure}[t]
\centerline{\psfig{figure=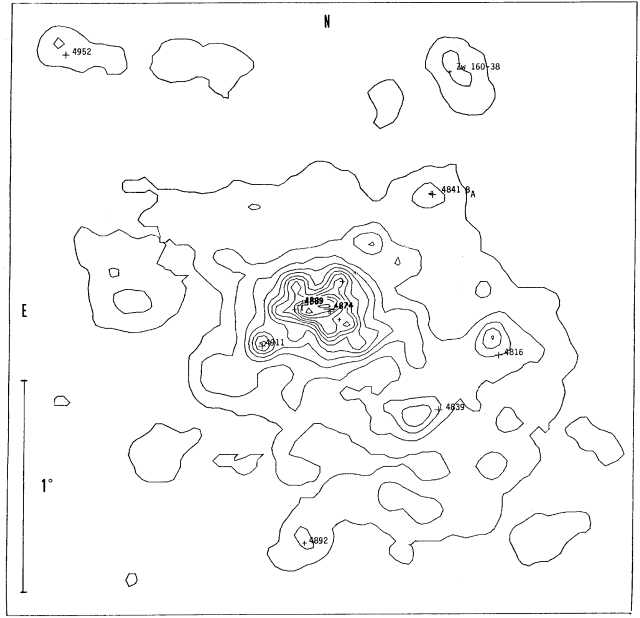,width=8.5cm,height=8.5cm}}
\caption{Isocontours of galaxy number densities in the Coma cluster; the 
brightest Coma galaxies are listed. The vertical line on the left indicates
a scale of 1$^{\circ}$ -- from Mellier et al.}
\label{f-mellier}
\end{figure}

The evidence for subclustering in the Coma core was growing fast. 
Perea et al. first identified the two central groups in velocity space,
and found mean velocities of 6431~km/s and 7072~km/s, and velocity dispersions
of 220~km/s and 137~km/s for the groups centered
around NGC~4889 and NGC~4874, respectively. Their results anticipated those of
Fitchett \& Webster and Mellier et al., both qualitatively and quantitatively
(Mellier et al.'s estimates of the velocity moments of the
two central groups differ by less than 10~\% from Perea et al.'s),
yet Perea et al.'s paper received much less attention 
than the other two\footnote{The {\em NASA Astrophysics Data
System} lists only three papers making reference to Perea et al.,
as compared to 75 papers referring to Fitchett \& Webster and 43 papers to
Mellier et al.} (was the word "taxonomical" in the title of their paper weird 
enough to discourage potential readers?).

Fitchett \& Webster were possibly the first to use the word "substructure" in
the title of a paper on Coma. Using the Lee-method, they separated galaxy
members of the NGC~4889 group from galaxy members of the 
NGC~4874 group, and proposed 
a circular-orbit dynamical model for the two central subgroups
(reminiscent of Valtonen \&
Byrd's model). They mentioned two possible origins for the substructures in the
core:
\begin{enumerate}
\item Coma is presently forming by the merger of two large 
subunits;
\item the two substructures in the Coma core were independent groups
that have fallen into a pre-existent cluster.
\end{enumerate}
Nowadays, a third interpretation\cite{cd} has made his way: 
\begin{enumerate}
\setcounter{enumi}{2}
\item NGC~4874 is the original dominant galaxy of the cluster, that has
recently suffered from a collision with a small group centered on
NGC~4889.
\end{enumerate}
The difference between the NGC~4874 velocity and the mean cluster velocity
(as well as its radio-morphology, see below),
still gives some credit to Fitchett \& Webster's scenario n.2
(see Biviano et al.\cite{b96}).

Of the 9 density peaks identified by Mellier et al. in Coma, 
two of them are in the cluster core (see Fig.\ref{f-mellier}), centered on
NGC~4889 and NGC~4874, that are surrounded by a
population of satellite galaxies. In Mellier et al.'s scenario, these 
subclusters exist as independent groups before infalling into the cluster.
At the time of infall, they have already passed
through a mass-segregation instability-phase.
During the infall, tidal effects strip the groups of their 
less bound low-mass galaxies, while the high-mass galaxies in the group
cores, resist longer. Support from their scenario came recently
from Biviano et al.\cite{b96}, who showed
that faint galaxies ($b \geq 17$) do not cluster around the giant
galaxies, at variance with bright galaxies. 

Mellier et al. estimated masses of $6 \times 10^{13} M_{\odot}$ 
and $5 \times 10^{13} M_{\odot}$, for the groups around
NGC~4874 and NGC~4889, respectively.
Their estimates have since been confirmed (to within 50~\%) by later works
(see below).

Some years later, Escalera et al.'s wavelet analysis 
detected the two central substructures with 99.9~\% significance. The
velocity dispersions they derived for the two subclusters are rather high,
and more similar to the values found by Fitchett \& Webster than to
those found by Perea et al. and Mellier et al. Probably, contamination by
cluster (and not group) members, affected their $\sigma_v$-estimates.

Non-optical wavelength observations also provided evidence for 
for a complex structure of the Coma core. In 1985, Feretti \&
Giovannini\cite{fg}'s radio observations of NGC~4874 showed that this galaxy
was a Wide-Angle-Tail radio-source. This morphology is indicative of motion
of NGC~4874 through the surrounding IC gas, i.e. the galaxy is not at rest 
at the bottom of the cluster potential (or the IC gas has a bulk velocity).

\begin{figure}
\centerline{\psfig{figure=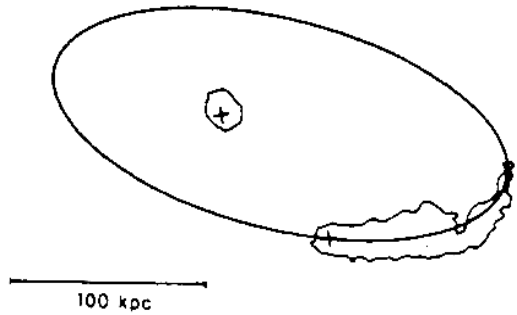,width=8cm,height=5.1cm}}
\caption{The orbit of NGC~4869 around NGC~4874 as derived by Feretti et
al. The radio-morphologies of the two galaxies are sketched.}
\label{f-feretti}
\end{figure}

Based on the radio morphology of another galaxy in the Coma core, NGC~4869, 
whose radio-emission was first mapped by Willson, 
Feretti et al.\cite{fdgv} obtained the orbit of this galaxy around NGC~4874
(see Fig.\ref{f-feretti}), and estimated the mass of the couple
in $5 \times 10^{13} M_{\odot}$ (similar to Mellier et al.'s estimate).

Last came the X-ray observations. In 1993 Davis \& Mushotzky\cite{dm} and
White et al. detected substructures in the X-ray image of Coma. In particular,
Davis \& Mushotzky\footnote{When Davis \& Mushotzky first found
evidence for substructure in the core of Coma {\em in the X-ray,} this
was known in the {\em optical} since 20 years\cite{bah2}.
In perspective, the following statement of them sounds auto-ironic:
\begin{quotation}
{\em The ability of the x-ray observations to locate this structure when both
galaxy counts and radial velocity information were inconclusive 
[\ldots] is an indication of the power of searching for substructure in x-rays.}
\end{quotation}}
used the {\em Einstein} data to show the existence of
excess X-ray emission from NGC~4874, and a region 2.5' west of NGC~4889. 
The higher resolution {\em ROSAT} data allowed White et al. to confirm
the former detection, but not the latter. Other irregularities in the
X-ray surface brightness were found, all associated with bright galaxies, 
NGC~4889, NGC~4911, NGC~4848 and NGC~4839 (this last already mentioned in
\S~\ref{ss-sw}). The groups detected by Mellier et al. in the optical were
showing up in X-rays!  

White et al. also detected a secondary peak of emission, half-way between the 
two central dominant galaxies, which they interpreted as gas stripped off 
NGC~4889. Biviano et al.\cite{b96} found an intriguing
positional coincidence of this X-ray peak and the number 
density peak of faint Coma galaxies. They suggested that White et
al.'s secondary peak is in fact emission from the main body of the Coma
cluster, masked by the superposition of several subclusters.

Finally, Vikhlinin et al.\cite{vfj} applied the wavelet analysis to the {\em
ROSAT} image of Coma, disentangling the emission of the two central groups
from that of the cluster. The optical\cite{esm} and X-ray wavelet-images 
looked remarkably similar. Under quite general assumptions, they
were able to estimate the masses of the NGC~4874 and NGC~4889 subclusters in
$3.4 \times 10^{13} M_{\odot}$ and $2.6 \times 10^{13} M_{\odot}$, again quite
close to the original Mellier et al.'s estimates.

\subsection{The Radio Halo Coma C}
A complete history of the Coma structure would not be complete without
mentioning the radio-halo of Coma. In fact, theory and observations suggest
that it is related to the subclustering in Coma.

Cluster radio-halos are very rare;
Hanisch\cite{han} searched for Coma-like radio-halos in 72 nearby Abell
clusters and did not find any! So, Coma is rather exceptional in this respect
(there are only $\sim 10$ cluster radio-halos known, see Feretti, these
proceedings). Its 
radio-halo, Coma C, was first detected at 408~MHz by Large et al.\cite{lmh}, 
as an extended source of 45' size at the Coma centre. Willson showed that Coma C
could not be produced by the integrated radiation from normal galaxies. Kim et
al.\cite{kkgv} and Venturi et al.\cite{vgf} 
found that Coma C extends to the SW, and
Giovannini et al.\cite{gfs} proposed the existence of a unique source extending
from Coma C to 1253$+$274, passing through the SW group around NGC~4839 (see
\S~\ref{ss-sw}). Recently, Feretti et al.\cite{fdgt} measured a magnetic field
associated with the cluster of $\sim 8.5 \mu$G, tangled on scales $\leq 1$~kpc.

The theorists had trouble in explaining the energy source of Coma C (see, e.g.,
Tribble\cite{tr1}). The radio-halo can be powered by relativistic electrons
moving in a magnetic field.
Cluster radio-galaxies can provide the relativistic electrons, 
but the strength of the magnetic field
and the large extent of the radio-halo imply that the electrons must
be re-accelerated far from their sources.

In the currently best model, recent ($\sim 10^8$ years)
subcluster collisions provide the re-acceleration energy 
(see Tribble\cite{tr2}). However, many clusters
contain substructures, and only a few clusters contain radio-halos, so
the situation is not so simple (see Feretti's
contribution in these proceedings for a discussion on this topic).

\section{An Old Vision of a New Cluster: Summary and Conclusions}
At the end of this long yet not exhaustive historical review, one is left with
the feeling that everything was already known since long ago. It is sufficient
to have a quick look to the important steps in our scientific understanding
of the Coma cluster, before 1980:
\begin{itemize}
\item{1901:} Wolf gives a map of Coma in which the SW group is already clearly
visible;
\item{1937:} Zwicky discovers the "missing-mass" problem;
\item{1954:} Shane \& Wirtanen's galaxy counts also show very clearly the SW
subcluster;
\item{1957:} Zwicky finds that bright and faint galaxies have different
radial distributions;
\item{1958:} Zwicky shows that the "missing-mass" problem is a "dark-matter"
problem, because clusters are stable and non-expanding;
\item{1959:} Abell finds the secondary peak in the otherwise monotonically
increasing luminosity function of Coma galaxies;
\item{1959:} Large et al. detect Coma C, the radio-halo, at 408~MHz;
\item{1960:} Mayall shows that the velocity dispersion decreases with
increasing clustercentric distance;
\item{1961:} van den Bergh makes the first objective detection of subclustering
in Coma;
\item{1966:} Reaves suggests that the lack of dwarf galaxies in the Coma core
is due to tidal disruption;
\item{1971:} Meekins et al. discover the X-ray emitting IC gas in Coma;
\item{1973:} Bahcall suggests the existence of subclustering around 
each of the two central dominant galaxies;
\item{1973:} des For\^ets \& Schneider show that galaxies of different types
have different velocity distributions;
\item{1975:} Lecar shows that the lack of significant luminosity
segregation imply that cluster galaxies have lost their halos;
\item{1978:} Sullivan \& Johnson find three HI-deficient spirals in Coma;
\item{1979:} Johnson et al.'s X-ray map of Coma hints at the presence of a SW
extension.
\end{itemize}

At the time they were produced, many of these early results needed firm
confirmation, that eventually came from more (and more accurate) data;
on the other hand, our theoretical understanding of these observational 
evidences is still far from complete. However, it seems to me that the
picture of a dynamically young Coma cluster was already contained in these
early results. So, maybe a more appropriate choice for this conference title
could have been: "An Old Vision of a New Cluster".

\section*{Acknowledgments}
This paper is dedicated to my wife Patrizia, for sharing my busy life of
wandering astronomer.

No historical review is unbiased. I apologize for the excessive emphasis I may
have put on my personal results, and for all other people's results that I have
misquoted or forgotten to mention.

I wish to thank Fabienne Casoli, Florence Durret, Daniel Gerbal, Alain Mazure, 
for the perfect organization of this conference. I thank Michael West for
pointing out to me Herschel's work on the Coma cluster.
I acknowledge the hospitality of the Trieste Astronomical Observatory, 
where a significant part of my bibliographic research was done.

\end{document}